\documentclass[a4paper,superscriptaddress,aps,prb,twocolumn,floatfix,citeautoscript,showpacs]{revtex4}

\usepackage{graphicx}
\usepackage{color}
\usepackage{amsmath}
\usepackage{amssymb}
\usepackage[english]{babel}
\usepackage[utf8]{inputenc}
\newcommand{\PRcomm}[1]{{\color{black} #1}}
\newcommand{\Pina}[1]{{\color{black} #1}}
\newcommand{\PinaR}[1]{{\color{black} #1}}

\usepackage{bm}
\usepackage{url}
\usepackage{ulem}
\newcommand{\ket}[1]{\left| #1 \right>} 
\newcommand{\bra}[1]{\left< #1 \right|} 
\graphicspath{{figures/}}
\begin{document}
%
%
\title{Reduced Density-Matrix Functional Theory: correlation and spectroscopy}
\newcommand{\lpt}{Laboratoire de Physique Th\'eorique, CNRS, IRSAMC, Universit\'e Toulouse III - Paul Sabatier, 118 Route de Narbonne, 31062 Toulouse Cedex, France}
\newcommand{\lcpq}{Laboratoire de Chimie et Physique Quantiques, IRSAMC, Universit\'e Toulouse III - Paul Sabatier, CNRS, 118 Route de Narbonne, 31062 Toulouse Cedex, France}
\newcommand{\lsi}{Laboratoire des Solides Irradi\'es, \'Ecole Polytechnique, CNRS, CEA-DSM, 91128 Palaiseau, France}
\newcommand{\etsf}{European Theoretical Spectroscopy Facility (ETSF)}

\affiliation{\lpt}
\affiliation{\lcpq}
\affiliation{\lsi}
\affiliation{\etsf}
\author{S. Di Sabatino}
\email[]{disabatino@irsamc.ups-tlse.fr}
\affiliation{\lpt}
\affiliation{\etsf}
\author{J.A. Berger}
\affiliation{\lcpq}
\affiliation{\etsf}
\author{L. Reining}
\affiliation{\lsi}
\affiliation{\etsf}
\author{P. Romaniello}
\affiliation{\lpt}
\affiliation{\etsf}
\pacs{71.10.-w,71.27.+a,31.15.V-,79.60.Bm}

\keywords{...}
\begin{abstract}
In this work we explore the performance of approximations to electron correlation in reduced density-matrix functional theory (RDMFT) and of approximations to the observables calculated within this theory. Our analysis focuses on the calculation of total energies, occupation numbers, removal/addition energies, and spectral functions. We use the exactly solvable Hubbard molecule at 1/4 and 1/2 filling as test systems. This allows us to analyze the underlying physics and to elucidate the origin of the observed trends. For comparison we also report the results of the $GW$ approximation, where the self-energy functional is approximated, but no further hypothesis are made concerning the approximations of the observables. In particular we focus on the atomic limit, where the two sites of the molecule are pulled apart and electrons localize on either site with equal probability\Pina{, unless a small perturbation is present}: this is the regime of strong electron correlation. In this limit, using the Hubbard molecule at 1/2 filling with or without a spin-symmetry-broken ground state, \Pina{allows us to explore how degeneracies and spin-symmetry breaking are treated in RDMFT}. We find that, within the used approximations, neither in RDMFT nor in $GW$ the signature of strong correlation are present in the \Pina{spin-singlet ground state, whereas both give the exact result for the spin-symmetry broken case}. Moreover we show how the spectroscopic properties change from one spin structure to the other. \PRcomm{Our findings \Pina{can be generalized to other situations, \PinaR{which} allows us to make connections to real materials and experiment.}}

\end{abstract}
\date{\today}
\maketitle
\section{Introduction}
Strongly correlated electron systems exhibit remarkable electronic and magnetic properties, such as metal-insulator transitions, half-metallicity, or unconventional superconductivity, which make them among the most attractive and versatile materials \PRcomm{(see e.g. Refs \onlinecite{Imada98,Koedderitzsch03,Norman11})}. Typically these materials have incompletely filled d- or f-electron shells with narrow energy bands. In this case a theoretical description based on current mean-field or perturbation approaches is not enough and a more accurate treatment of electron correlation is required \PRcomm{\cite{gatti-VO2,Aryasetiawan-Tmatrix,Aryasetiawan95,Biermann03}}. This represents one of the greatest challenges in condensed-matter physics today.

One of the most popular approaches in condensed-matter physics is many-body perturbation theory (MBPT) based on Green's functions. Within the so called $GW$ approximation \cite{hedin65} to electron correlation, MBPT has become, over the last two decades, the tool of choice for the calculations of quasiparticle (QP) band structures \cite{Aulbur,Gunnarsson,vidal2010,zunger2011,chulkov2013,gonze2013} and direct and inverse photo-emission spectra \cite{gatti-V2O3,Kotani07,gatti-VO2,Kotani04,zunger2012} of many materials improving substantially over the results provided by static mean-field electronic structure methods. However $GW$ suffers from some fundamental shortcomings  \cite{Godby,pina09,vanSchilfgaarde06,Stan09,Caruso12,vonFriesen09}, and, in particular, it is not expected to describe strong correlation. More refined levels of approximations are hence needed and much effort is devoted to this goal both by going beyond standard methods \cite{Aryasetiawan-Tmatrix,chulkov2004,romaniello_PRB12, guzzo_prl,guzzo2014,louie2013,kresse2014} and by exploring novel routes to calculate Green's functions \cite{lani_njp,berger_njp}.  In this context, promising results for solids have been reported using reduced density-matrix functional theory (RDMFT) \cite{Sharma13}. 

Within RDMFT the ground-state properties of a physical system are functionals of the ground-state density matrix \cite{PhysRev.97.1474, Gilbert75}, since there exists a one-to-one mapping between the (non-degenerate) ground-state wavefunction of the system and the corresponding density matrix  \cite{Gilbert75}. In particular the ground-state total energy is a functional of 
the one-body density matrix $\gamma$ and it can be written as $E_{tot}[\gamma]=E_{kin}[\gamma]+E_{ext}[\gamma] + E_{Hxc}[\gamma] $, where $E_{kin}$, $E_{ext}$, and $E_{Hxc}$, are the kinetic energy, the energy due to the coupling to an external potential, and the Hartree and exchange-correlation energies, respectively. Energy minimization under the constraint that $\gamma$ is $N$-representable, determines the exact $\gamma$. In practice, however, approximations to 
$E_{xc}[\gamma]$ are needed. Several approximations have been proposed and most of them are implicit functionals of the density matrix; they are explicit functionals of the natural orbitals \cite{PhysRev.97.1474}  $\phi_i$ and occupation numbers $n_i$, i.e. the eigenvectors and eigenvalues, respectively, of the density matrix, $\gamma(\mathbf{x},\mathbf{x}')=\sum_in_i\phi_i(\mathbf{x})\phi^*_i(\mathbf{x}')$.  The total energy is then a functional of $\phi_i$ and $n_i$.
Once the density matrix of the system is known, all the observables of the system can be calculated, provided that their expression as functional of the density matrix is known. 

This is not the case for the one-body spectral function, which determines, for example, photoemission spectra. Various ways to calculate removal/addition energies have been proposed \cite{Pernal200571,Sharma13}. For example, removal energies can be calculated by using \PRcomm{the method proposed by Pernal and Cioslowski \cite{Pernal200571}, which is based on the extended Koopmans' theorem (EKT)  \cite{Day74,Morrell75}}. So far, the method has been used only for finite systems. \PRcomm{Numerical evidence suggests that EKT is exact only for the lowest ionization potential \cite{Morrison92,Sundholm93}}. In Ref.\ \onlinecite{Sharma13} an approximate procedure to calculate quasiparticle energies and photoemission spectra within RDMFT has been proposed, which is also inspired by Koopmans' theorem. When applied to a series of transition-metal oxides, the method seems to capture the essential physics of strong electron correlations. \Pina{These are, however, only empirical evidences, and an in-depth analysis is missing. This is not simple because }several approximations are involved: i) an approximate exchange-correlation energy functional, ii) an approximate expression for the removal and addition energies, iii) an approximate expression for the spectral function.
\Pina{\PinaR{It is therefore important} to study these aspects in a systematic way \PinaR{in order to advance our understanding of an approach which is used all over physics and chemistry}. \PinaR{To do this, we need a simpler system, preferably with a known exact solution for benchmarking, with a direct link between the molecular orbitals and natural orbitals, with the possibility to study quasi-degeneracies and (spin and charge) symmetry breaking. An ideal candidate is the Hubbard dimer}: it is exactly solvable, the natural orbitals correspond to the bonding/antibonding orbitals, and the atomic limit $t\rightarrow 0$ offers a playground to explore degeneracies and symmetry breaking.  In the atomic limit, indeed, when the two sites are pulled apart, electrons localize on either site with equal probability, unless a small perturbation is present: this is the regime of strong electron correlation. In this limit, all the eigenstates of the system at 1/4 filling acquire equal energy and they become degenerate with the charge-symmetry broken states; at 1/2 filling the spin-singlet ground state becomes degenerate with the spin-triplet state as well as with the spin-symmetry-broken states (see Tables I and II in Ref.\ \onlinecite{pina09}). This scenario is general and common also to other molecules, such as, e.g. $H_2$ at dissociation\PinaR{, which is a paradigmatic example in quantum chemistry} (see, e.g. \onlinecite{Baerends01,Fuchs05,Wu07,Giesbertz08,Dahlen06,Stan06,Stan09,Ren12,Caruso13,Olsen14}). Analogies can be found also in infinite systems, as, for example, in the homogeneous electron gas (HEG). In the HEG the analogous of the bonding/antibonding orbitals  are the eigenstates of the perfectly translationally invariant system, which are also the natural orbitals. \PinaR{At low densities electrons localize to minimize the electron-electron interaction and the translational symmetry is spontaneously broken.}

In the following, therefore, we will use the Hubbard dimer at 1/4 and 1/2 filling as a test case, \PinaR{suggesting extrapolation to real systems when appropriate.}}
The paper is organized as 
follows. In Sec.\ \ref{Sec:Theory} we will report the key equations of MBPT and RDMFT. In Sec.\ \ref{Sec:Results} MBPT and RDMFT results for occupation numbers, total energies, removal/addition energies, and spectral function are compared to exact results and analyzed. Conclusions are given in Sec.\ \ref{Sec:Conclusions}
\section{Theoretical framework\label{Sec:Theory}}
In the following we will give the key equations used in many-body perturbation theory and reduced density-matrix functional theory, and in particular we will discuss how one can calculate ground- and excited-state properties, namely total energies, occupation numbers, removal and addition energies, and spectral function in the two approaches. We will use atomic units $\hbar=m=e=1$ and work at zero temperature throughout the paper .


\subsection{MBPT}
Within MBPT the leading role is played by the one-body Green's function. At zero temperature the time-ordered \PinaR{equilibrium} Green's function \cite{fetterwal} reads
\begin{equation}
G(1,2)\equiv-i\bra{\Psi_0}\mathcal{T}[\hat{\psi}(1)\hat{\psi}^\dagger(2)]\ket{\Psi_0},
\label{Eqn:G}
\end{equation}
where $\mathcal{T}$ is the time-ordering operator, $\Psi_0$ is the ground-state many-body wavefunction, $\hat{\psi}$ and $\hat{\psi}^\dagger$ are field operators in the Heisenberg picture. Here $(1)\equiv\displaystyle  (\mathbf{x}_1,t_1)\equiv \displaystyle  (\mathbf{r}_1,s_1,t_1) $, $(1^+)\equiv(\mathbf{x}_1,t^+_1)$ with $t_1^+\equiv t_1+\delta$ ($\delta\rightarrow 0^+$) describe space, spin and time coordinates. $G$ contains a wealth of information about a physical 
system. In particular the ground-state total energy can be obtained using the Galitskii-Migdal formula
\begin{equation}
E_0=-\frac{i}{2}\int d\mathbf{x}_1\lim_{2\rightarrow 1^+}\left(i\frac{\partial}{\partial t_1}+h_0(\mathbf{r}_1)\right)G(1,2)
\label{Eqn:GM}
\end{equation}
where 
$h_0(\mathbf{r}_1)=-\nabla_{\mathbf{r}_1}/2+v_{ext}(\mathbf{r}_1)$ is the one-body hamiltonian. Moreover we are also interested in the density matrix $\gamma(\mathbf{x}_1,\mathbf{x}_2)=-iG(1,\mathbf{x}_2 t_1^+)=-i\int d\omega/(2\pi)\, G(\omega)e^{i\omega0^+}$, and the spectral function $A(\omega)=|\mathcal{\Im}G(\omega)|/\pi$, which is closely related to photoemission spectra \cite{hedin99}. 

Equation (\ref{Eqn:G}) is not practical to determine $G$, since it requires the knowledge of the many-body wavefunction. In MBPT one uses instead the Dyson equation $G=G_0+G_0\Sigma G$, where $G_0$ is the non-interacting one-body Green's function and $\Sigma$ is the self-energy, which describes all the many-body effects of the system. Approximations to the self-energy are needed, and a very popular one is $\Sigma=v_H+iGW$, where $v_H$ is the Hartree potential and $W=\epsilon^{-1}v_c$ is the dynamically screened Coulomb interaction, with $\epsilon^{-1}$ the inverse dielectric function and $v_c$ the Coulomb interaction \cite{hedin65}. In Sec.\ \ref{Sec:Results} we will compare the exact results obtained using (\ref{Eqn:G}) with those obtained from the $GW$ approximation. The $GW$ equations should, in principle, be solved self-consistently, since the self-energy is a functional of the one-body Green's function. Here, instead, we solve the $GW$ equations without self-consistency, which is often the case in practice; \PRcomm{we calculate $G=[1-G_0\Sigma]^{-1}G_0$ and \Pina{we use $\Sigma=v_H+iG_0W_0$, with $v_H=-iv_cG_0$ and $W_0=[1+iv_cG_0G_0]^{-1}v_c$.}
}

\subsection{RDMFT}
The reduced \textit{p}-body density matrix is defined as
\begin{multline*}
\Gamma^{(p)}(\mathbf{x}_1...\mathbf{x}_p,\mathbf{x}_1'...\mathbf{x}_p')\equiv\\
\binom{N}{p}\int d \mathbf{x}_{p+1}...d\mathbf{x}_N\Psi^*(\mathbf{x}_1'...\mathbf{x}'_p,\mathbf{x}_{p+1}...\mathbf{x}_N)\\
\times\Psi(\mathbf{x}_1...\mathbf{x}_p,\mathbf{x}_{p+1}...\mathbf{x}_N)
\end{multline*}
Within RDMFT \cite{Gilbert75} the total energy is a unique functional of the one-body density matrix $\gamma\equiv\Gamma^{(1)}$. It reads
\begin{eqnarray*}
E_{tot}[\gamma]&=&E_{kin}[\gamma]+E_{ext}[\gamma]\\
&&+\int \! d \mathbf{x}\, d \mathbf{x}^\prime \, v_c(\mathbf{x},\mathbf{x}^\prime)
\Gamma^{(2)}[\gamma](\mathbf{x},\mathbf{x}^\prime ; \mathbf{x},\mathbf{x}^\prime) 
\end{eqnarray*}
with
\begin{equation}
 \Gamma^{(2)}[\gamma](\mathbf{x},\mathbf{x}^\prime; \mathbf{x},\mathbf{x}^\prime)=
\frac{1}{2} \gamma(\mathbf{x},\mathbf{x})
 \gamma(\mathbf{x}^\prime, \mathbf{x}^\prime)
 +\Gamma^{(2)}_{xc}[\gamma](\mathbf{x},\mathbf{x}^\prime; \mathbf{x}, \mathbf{x}^\prime).
 \label{Eqn:Gamma_2}
\end{equation}
The first and second terms on the right-hand side of (\ref{Eqn:Gamma_2}) are the Hartree and exchange-correlation contributions, respectively. The latter is not known and needs to be approximated. Various approximations have been proposed in literature \cite{Gritsenko,BuijsePhD,Buijse2,GU_PRL98,Lathiotakis_PRB07,Piris_IJQC13,sharma_PRB08}. Many of them, however, can be traced back to the work of M\"uller \cite{muller}, and are based on the factorization
\begin{equation}
  \Gamma_{xc}^{(2)}[\gamma](\mathbf{x},\mathbf{x}^\prime;\mathbf{x},\mathbf{x}^\prime)\approx-\frac{1}{2}\gamma^\alpha(\mathbf{x}^\prime, 
  \mathbf{x})\gamma^{\alpha}(\mathbf{x}, \mathbf{x}^\prime)
  \label{Eqn:xc_approx}
\end{equation}
where
\begin{equation*}
 \gamma^\alpha(\mathbf{x},\mathbf{x}^\prime)=\sum_{j}^{}n^\alpha_{j}\phi_{j}(\mathbf{x})\phi^*_j(\mathbf{x}^\prime)
\end{equation*}
wihere $\phi_{i}(\mathbf{x})$ and $n_{i}$ are the naturals orbitals and the occupation numbers, respectively. Note that with $\alpha=1$ in (\ref{Eqn:xc_approx}) one gets the Hartree-Fock approximation to $\Gamma^{(2)}$. The total energy can then be expressed as a functional of $\phi_i$ and $n_i$, $E_{tot}[\{n_i\},\{\phi_i\}]$; 
functional minimization with respect to the natural orbitals, under orthonormality constraints, and occupation numbers, under total particle conservation and N-representability constraints ($0\leq n_i\leq 1$), leads to the ground-state total energy. 
\PRcomm{ In this work we study the approximation (\ref{Eqn:xc_approx}), with $0.5\leq \alpha\leq1$, which has been applied both to extended \cite{Lathiotakis_PRB07,sharma_PRB08,Lathiotakis_PRA09,Sharma13} as well as to finite systems \Pina{(see, e. g. Refs \onlinecite{Gritsenko,Pernal200571,Cohen_CPL02,Lathiotakis_JCP08,Lathiotakis_PRA09,Lathiotakis_JCP09}})}. From the natural orbitals and occupation numbers which minimize the total energy one can build the one-body density matrix \PinaR{that} corresponds to a given approximation for $\Gamma^{(2)}$. The procedure to determine the one-body density matrix is then different  from the standard one used for the Green's function, for which one solves a Dyson equation for a given approximation to the self-energy. Moreover, whereas one can easily extract information about photoemission spectra directly from the imaginary part of $G$, this is not the case in RDMFT. \Pina{It does not yield} $G$ nor its imaginary part.

Natural occupation numbers are strictly related to the multi-determinant nature of the wavefunction of a physical system. Let us expand the many-body wavefunction in terms of Slater determinants constructed from the eigenvectors of the one-body density matrix $\{\phi_i\}$, $\Psi_0(\mathbf{x}_1...\mathbf{x}_N)=\sum_i C_i \Phi_i(\mathbf{x}_1...\mathbf{x}_N)$.  The one-body density-matrix then reads \cite{giesbertz2010}
\begin{multline}
\gamma(\mathbf{x},\mathbf{x}')=\\
N\int d\mathbf{x}_2...d\mathbf{x}_N\sum_{ij}C^*_iC_j \Phi^*_i(\mathbf{x}',\mathbf{x}_2...\mathbf{x}_N) \Phi_j(\mathbf{x},\mathbf{x}_2...\mathbf{x}_N)\\
=\sum_{i}|C_i|^2\gamma_i(\mathbf{x},\mathbf{x}')
\label{Eqn:DM}
\end{multline}
where $\gamma_i(\mathbf{x},\mathbf{x}')=\sum_k\phi^i_k(\mathbf{x})\phi^{i*}_k(\mathbf{x}')$ is the density matrix associated to the $i$th Slater determinant. If the wavefucntion of the system is described by a single Slater determinant, as in the case of a single (spin-polarized) electron (see Hubbard molecule at 1/4 filling), then the natural occupation numbers are either 1 or 0. If instead more determinants are involved, the natural occupation numbers, in general, take fractional values between 0 and 1. This can be nicely illustrated by considering a two-electron system with a  singlet wavefunction $\Psi_0(\mathbf{x}_1,\mathbf{x}_2)=\sum_{i=1,2}C_i\Phi_i(\mathbf{x}_1,\mathbf{x}_2)$, where $\Phi_1=\ket{b\uparrow,b\downarrow}$ and $\Phi_2=\ket{a\uparrow,a\downarrow}$ are Slater determinants constructed from bonding and antibonding orbitals \PRcomm{$\{\phi_i\}$}, respectively (see Hubbard molecule at 1/2 filling). \Pina{Note that the bonding/antibonding orbitals in the Hubbard molecule correspond to the natural orbitals.} The one-body density matrix reads:
\begin{eqnarray}
\gamma(\mathbf{x},\mathbf{x}')&=&|C_1|^2\sum_{i=b\uparrow,b\downarrow}\phi_i(\mathbf{x})\phi^*_{i}(\mathbf{x}')\nonumber\\
&&+|C_2|^2\sum_{i=a\uparrow,a\downarrow}\phi_{i}(\mathbf{x})\phi^*_{i}(\mathbf{x}')\nonumber\\
&=&\sum_{i=b\uparrow,b\downarrow,a\uparrow,a\downarrow}n_{i} \phi_{i}(\mathbf{x})\phi^*_i(\mathbf{x}')
\end{eqnarray}
with $n_{b\uparrow}=n_{b\downarrow}=|C_1|^2$ and $n_{a\uparrow}=n_{a\downarrow}=|C_2|^2$, and $|C_1|^2+|C_2|^2=1$ since the wavefunction $\Psi_0$ is normalized. In general the relation between $C_i$ and natural occupation numbers is more complicated than in this example, but the fact that fractional occupation numbers reflect the multideterminant nature of the wavefunction, and hence the degree of correlation in a system, remains still valid.

Recently Sharma \textit{et al.} \cite{Sharma13} proposed the following approximate expression for the spectral function within RDMFT 
\begin{equation}
A(\omega)\approx \sum_i\left[n_i\delta(\omega-\epsilon_i^-)+(1-n_i)\delta(\omega+\epsilon_i^+)\right]
\label{Eqn:DOS}
\end{equation}
where $\epsilon_i^{\pm}=E^N_0-E^{N\pm 1}_i$, with $E^N_0$ \PinaR{the ground-state energy of the $N$-electron system and $E^{N\pm 1}_i$ the ith-state energy of the $N\pm 1$-electron system.} To arrive at Eq. (\ref{Eqn:DOS}) one starts from the exact expression $A(\omega)=|\Im G(\omega)|/\pi$ and approximates the ground and excited states of the $N+1$- and $N-1$-electron systems by adding an electron, $|\Psi^{N+1}_i\rangle=\frac{1}{\sqrt{1-n_i}}c^\dagger_i|\Psi^{N}_0\rangle$, or a hole, $|\Psi^{N-1}_i\rangle=\frac{1}{\sqrt{n_i}}c_i|\Psi^{N}_0\rangle$, to the ground state of the N-electron system. This is in the spirit of Koopmans' theorem and it is an approximation, because in general the set of states obtained in this way are not eigenstates of the $N+1$- and $N-1$-electron system, respectively, and do not form a complete set. Along the same line, 
the energies $\epsilon_i^-$ and $\epsilon_i^+$ in Eq. (\ref{Eqn:DOS}) are calculated in an approximate way as 
\begin{equation}
\epsilon^{-}_k=-\epsilon^{+}_k=\epsilon_k=\left .E[\{n_i\},\{\phi_i\}]\right|_{n_k=1}-\left .E[\{n_i\},\{\phi_i\}]\right|_{n_k=0},
\label{Eqn:DIF}
\end{equation} 
where $\left .E[\{n_i\},\{\phi_i\}]\right|_{n_k=1}$ ($\left .E[\{n_i\},\{\phi_i\}]\right|_{n_k=0}$) is the total energy for the $N$-particle system with all the occupation numbers fixed at their optimal value (i.e. the value that minimize the energy functional) except for the occupation number $n_k$ which is fixed to 1 (0). We will refer to this method as DIF, to keep contact with other works on the subject \cite{PhysRevA.85.032504}. Using (\ref{Eqn:DIF}) for the calculation of removal and addition energies the expression of the spectral function in (\ref{Eqn:DOS}) simplifies to $A(\omega)=\sum_i\delta(\omega-\epsilon_i)$.

We note that the energies calculated using Eq. (\ref{Eqn:DIF}) have both a removal and addition character, because, in general, the state $k$ is partially filled. Equation (\ref{Eqn:DIF}) can, indeed, be rewritten as the sum of two contributions
\begin{eqnarray}
\epsilon_k&=&\left(\left.E[\{n_i\},\{\phi_i\}]\right|_{n_k=1}-\left.E[\{n_i\},\{\phi_i\}]\right|_{n_k=n_k^{opt}}\right)\nonumber\\
&+&\left(\left.E[\{n_i\},\{\phi_i\}]\right|_{n_k=n_k^{opt}}-\left.E[\{n_i\},\{\phi_i\}]\right|_{n_k=0}\right)
\label{Eqn:Analyse}
\end{eqnarray}
where $n_k^{opt}$ are the occupations numbers which minimize the total energy. The first term on the right-hand side of Eq.\ (\ref{Eqn:Analyse}) corresponds to the addition of a fraction of electron equal to $1-n_k^{opt}$ while the second to the removal of a fraction of electron equal to $n_k^{opt}$.

Moreover the number of energies calculated using (\ref{Eqn:DIF}) equals the number of occupation numbers (i.e., the dimension of the natural orbital basis set), which is in general smaller than the exact number of removal and addition energies; it equals, indeed, the number of noninteracting states and hence the number of quasiparticles. \Pina{Note that quasiparticle peaks in the spectral function can be directly linked to peaks in the non-interacting spectral function, whereas satellites are additional structures which are generated by the frequency-dependence of the self-energy and, therefore, have zero spectral weight for vanishing interaction. The spectral weight of a quasiparticle peak, instead, remains constant or might decrease by increasing the interaction, the weight being transferred to the satellites.} This can be illustrated using the Hubbard molecule. As shown in Sec.\ {\ref{Sec:Results} for this model system the basis of natural orbitals $\{\phi_i\}$ diagonalizes also the one-body Green's function for any frequency, therefore one can write 
\begin{equation}
G(\mathbf{x}_1,\mathbf{x}_2;\omega)=\sum_iG_i(\omega)\phi_i(\mathbf{x}_1)\phi^*_i(\mathbf{x}_2),
\end{equation}
and for the occupation numbers one gets
\begin{equation}
n_i=-i\int\frac{d\omega}{2\pi}G_i(\omega)e^{i\omega0^+}.
\label{Eqn:N-G}
\end{equation}
 If $G_i$ has more than one pole, than the total number of removal/addition energies that one should find is larger than the number of occupation numbers. Therefore, equation (\ref{Eqn:DIF}), in general, describes a mixture of quasiparticle and satellite energies, as will be illustrated in Sec.\ {\ref{Sec:Results}.
 
 The total energy difference, Eq.\ (\ref{Eqn:DIF}),  can be further approximated as 
\begin{equation}
\left .E[\{n_i\},\{\phi_i\}]\right|_{n_k=1}-\left .E[\{n_i\},\{\phi_i\}]\right|_{n_k=0}\approx \left.\frac{\partial E}{\partial n_k}\right|_{n_k=1/2}
\label {Eqn:DER}
\end{equation} 
which is justified if the total energy is linear in the occupation number $n_k$ \cite{Liberman00}. This method will be referred to as DER. 
Using Eqs (\ref{Eqn:DOS}) and (\ref{Eqn:DER}) the spectral function of several transition metal oxides 
has been calculated, showing that some experimental features are captured \cite{Sharma13}.

As an alternative to Eq. (\ref{Eqn:DIF}) (or (\ref{Eqn:DER})) removal energies can be calculated by using extended Koopmans' theorem (EKT) as proposed by Pernal and Cioslowski \cite{Pernal200571}. The method is based on the diagonalization of the Lagrangian matrix :
\begin{equation}
\Lambda_{ij}=\frac{1}{\sqrt{n_in_j}}\left[n_ih_{0,ji}+2\sum_{klm}\Gamma^{(2)}_{iklm}v_{c,jkml}\right]
\label{Eqn:Lagrangian}
\end{equation}
with $h_{0,ji}=\int d\mathbf{x}\,\phi_j^*(\mathbf{x})h_0(\mathbf{x})\phi_i (\mathbf{x})$, 
\begin{eqnarray*}
 \Gamma^{(2)}_{iklm}&=&\int d\mathbf{x}'_1d\mathbf{x}'_2d\mathbf{x}_1d\mathbf{x}_2\,\Gamma^{(2)}(\mathbf{x}'_1,\mathbf{x}'_2;\mathbf{x}_1,\mathbf{x}_2)\\
 &&\times\phi_m^*(\mathbf{x}_2)\phi_l^*(\mathbf{x}_1) \phi_k(\mathbf{x}'_2)\phi_i(\mathbf{x}'_1),
\end{eqnarray*}
and 
\begin{eqnarray*}
 v_{c,jkml}=\int d \mathbf{x}_1d\mathbf{x}_2 \phi_j^*(\mathbf{x_1}) \phi_k^*(\mathbf{x_2})v_c(\mathbf{x_1},\mathbf{x_2})\phi_m(\mathbf{x_1}) \phi_l(\mathbf{x_2}).
\end{eqnarray*}
 The eigenvalues of $\Lambda$ are the removal energies. The underlying physics of this method is similar to that of Eq.(\ref{Eqn:DOS}), although more advanced: in the EKT the $N-1$-electron states are obtained as a linear combination of states obtained by removing an electron from the ground state of the $N$-electron system, $|\Psi^{N-1}\rangle=\sum_iB_ic_i|\Psi_0^{N}\rangle$; the energy of the so obtained $N-1$-electron states is minimized with respect to the coefficients $B_i$, unlike in the DIF/DER method. In practice the EKT has only been applied to finite systems. Although it is an approximate method for the calculation of removal energies, for the Hubbard molecule at 1/4 and 1/2 filling it delivers the exact results when combined with the exact exchange-correlation energy functional. Therefore, in this work we will use it to test approximations to the xc energy functional. Note that the lowest addition energy can be obtained from the highest removal energy of the $N+1$-system (if the latter is stable).

%

%
%
%
\section{Correlation in the Hubbard molecule\label{Sec:Results}}
In this section we will illustrate the physics behind different approximations to correlation as well as to observables in RDMFT and show how it compares with the standard $GW$ method used in MBPT. To this purpose we use the Hubbard molecule, a simple prototype of a strongly correlated system that can be solved exactly \cite{Hubbard26111963}. The Hamiltonian of the Hubbard molecule reads
\begin{multline}
 H = -t \sum_{\substack{i,j=1,2\\i\neq j}}\sum_{\sigma} c^{\dagger}_{i,\sigma} c_{j,\sigma}
 \\+\frac{U}{2}
 \sum_{i=1,2}\sum_{\sigma\sigma^\prime}c^\dagger_{i,\sigma}c^\dagger_{i,\sigma^\prime}c_{i,\sigma^\prime}c_{i,\sigma} + \epsilon_0 
 \sum_{\sigma,i=1,2} c^{\dagger}_{i,\sigma}c_{i,\sigma}+ V_0.
 \end{multline}
Here $c^{\dagger}_{i,\sigma}$ and  $c_{i,\sigma}$ are the creation and annihilation operators for an electron at site $i$ with spin $\sigma$, $U$ is the on-site (spin-independent) interaction, $-t$ is the hopping kinetic energy, and $\epsilon_0$ is the orbital energy. The Hamiltonian further contains a potential $V_0$ that can be chosen to fix the zero-energy scale. The physics of the Hubbard model \cite{Hubbard26111963} arises from the competition between the hopping term, which prefers to delocalize electrons, and the on-site interaction, which favours localization. The ratio $U/t$ is a measure for the relative contribution of both terms and is the intrinsic, dimensionless coupling constant of the Hubbard model, which will be used in the following.

We refer to Ref \onlinecite{pina09} and \onlinecite{romaniello_PRB12} for the exact results of the model at 1/4 and 1/2 filling, respectively. Here, however, instead of the site 
basis, which can be considered as an atomic-like basis set, we will use the bonding/antibonding basis set (see App. \ref{app_bondant}), which is conceptually similar to a molecular-like basis set. The reason behind this choice is that within this basis set the density matrix is diagonal; in other words this basis set is the basis of natural orbitals. 
\subsection{Total energy and occupation numbers}

\begin{figure}[t]
\begin{center}
\includegraphics{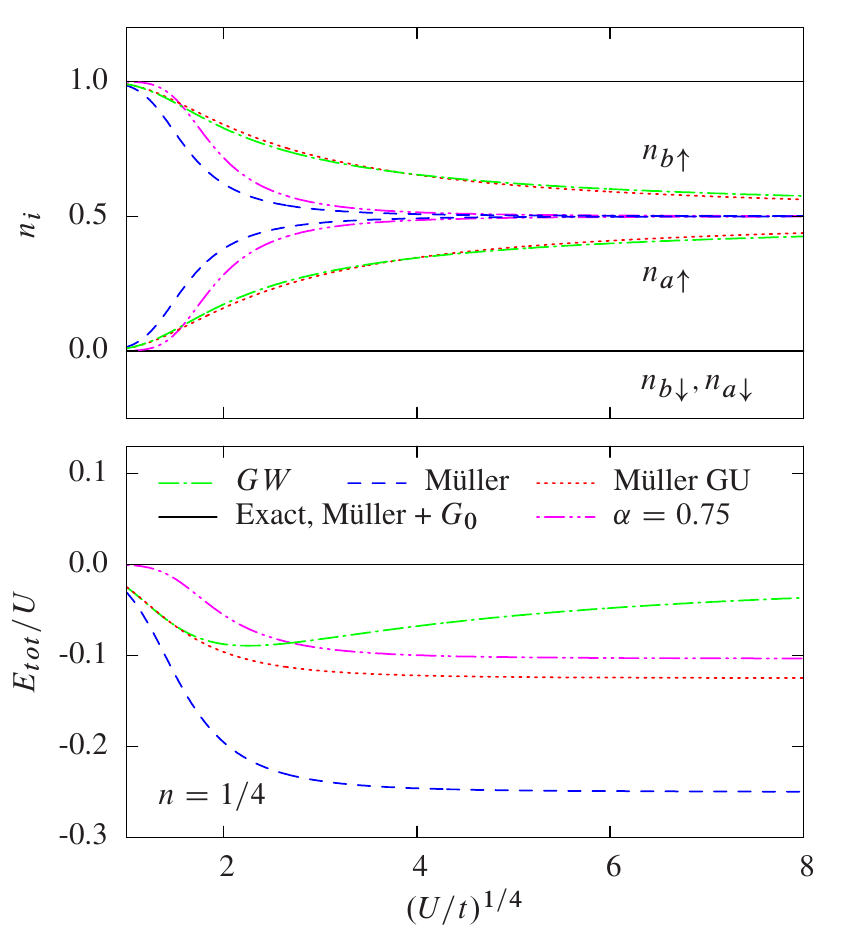}
\end{center}
\vspace{-15pt}
\caption{(Color online) Occupation numbers (upper panel) and total energy (lower panel) as function of $U/t$ at 1/4 filling: exact vs M\"uller functional w/wo self-interaction corrections (M\"uller and M\"uller GU, respectively) and $GW$. \Pina{Results obtained using $\alpha=0.75$ in the M\"uller-like xc funcional are also reported to show how the occupation numbers and the total energy vary by changing $\alpha$ between 0.5 and 1.} Total energies obtained using $G_0$ occupation numbers in the M\"uller functional and the $GW$ Green's function in the Galitskii-Migdal formula (labelled M\"uller+$G_0$ and $GW$, respectively) are also reported.}
\label{Fig:figure1}
\end{figure}

\subsubsection{1/4 filling\label{Sec:TOT_E_oneforth}}
In the case of the Hubbard molecule at 1/4 filling the ground-state wavefunction reads 
$\ket{\Psi_0}=\left(\ket{\uparrow\,0}+\ket{0\,\uparrow}\right)/\sqrt{2}$ in the site basis. When projected in the bonding/antibonding basis one gets $\ket{\Psi_0}=\ket{b\uparrow}$, i.e one single Slater determinant with one electron in the bonding orbital. The exact density matrix is idempotent for any $U/t$ value at zero temperature (see Fig.\ \ref{Fig:figure1} \Pina{\footnote{Note that here and in Fig. \ref{Fig:figure2} we use $(U/t)^{1/4}$ on the horizontal axis to facilitate the comparison with Ref.\ \onlinecite{Olsen14} at 1/2 filling.}}, upper panel), with the occupation number of the bonding (antibonding) orbital $n_{b\uparrow}=1$ and $n_{b\downarrow}=0$ ($n_{a\uparrow}=n_{a\downarrow}=0$). In particular in the atomic limit $t\rightarrow0$, for which $U/t\rightarrow \infty$ for $U$ fixed, the electron remains in the bonding orbital. Analyzed in the site basis this means that the electron has the same probability of being on either of the two sites. \PinaR{Often the electronic structure is probed by electron addition or removal, like in inverse or direct photoemission experiments. Since the system electron is in one of the two states, there are two possible addition energies:} one at $\epsilon_0$ when the addition electron goes to the unoccupied site, and one at $\epsilon_0 + U$ when it goes to the
occupied one. This is an example of strong correlation, since the added electron needs to see the electron in the system and not just an average charge distribution. The occupation numbers \PinaR{of the original system before the measurement are, instead, simply zero or one, because the initial state has no correlation.}
When we use the M\"uller functional  ($\alpha=0.5$ in Eq.\ (\ref{Eqn:xc_approx})), the optimal occupation numbers $n_{b\uparrow}$ and $n_{a\uparrow}$ tend to $0.5$ with increasing $U/t$ (see Fig.\ \ref{Fig:figure1}, upper panel). By increasing $\alpha$ up to 1 (Hartree-Fock) one approaches the  
exact situation,  \PRcomm{with the occupation numbers varying continuously with respect to $\alpha$ for each value of the interaction \Pina{(see, as an example, the results at $\alpha=0.75$ in Fig.\ \ref{Fig:figure1}})}. This is because exchange and Hartree energy completely cancel each other in the case of one electron. Self-interaction corrections introduced by Goedecker and Umrigar (GU) \cite{GU_PRL98} to the M\"uller functional (M\"uller GU) slow down the eventual merging of bonding and 
antibonding occupation numbers at 1/2. Interestingly the occupation numbers calculated from the $G_{GW}$ are almost on top of the self-interaction corrected M\"uller functional results. To obtain $GW$ occupation numbers we use Eq.\ (\ref{Eqn:N-G}), with the $GW$ one-body Green's function on the right-hand side.

Total energy results are reported in the lower panel of Fig.\ \ref{Fig:figure1}.  
All the approximations used underestimate the total energy, the exception being the M\"uller functional when fed with exact occupation numbers (obtained from $G_0$ at 1/4 filling\footnote{For one electron the hole part of the non-interacting Green's function, i.e. the part related to removal energies, is exact, because the removed electron does not interact with other electrons in the system; the exact occupation numbers can hence be calculated from $G_0$ according to Eq.\ (\ref{Eqn:N-G}). 
}), which is on top of the exact result (``M\"uller+$G_0$" in Fig.\ \ref{Fig:figure1}). This finding is also observed at 1/2 filling. However, at one fourth filling, the exact occupation numbers  being 1 or 0, any $0\leq \alpha \leq 1$ would give the exact total energy. 
For comparison we also reported the total energy obtained using the Galitskii-Migdal equation (\ref{Eqn:GM}) with the $GW$ Green's function, which is similar to the results obtained using the M\"uller GU functional \PRcomm{below $(U/t)^{1/4}\simeq 2$. In the atomic limit, however, the $GW$ total energy tends to the exact one,} as was already noticed in Ref. \cite{pina09}, whereas the M\"uller GU gives a lower energy. 

In conclusion at one fourth filling we find that the Hartree-Fock approximation ($\alpha=1$ in (\ref{Eqn:xc_approx})) gives the exact occupation numbers, whereas  the M\"uller functional ($\alpha=0.5$ in (\ref{Eqn:xc_approx})) gives results which quickly depart from the exact ones. Varying $\alpha$ between 0.5 and 1 or using self-interaction corrected M\"uller functional gives results in between these two extremes for any $U/t$; $GW$ occupation numbers are almost on top of self-interaction corrected M\"uller functional results. Concerning the total energy all M\"uller-like functionals fed with exact occupation numbers give the exact results. All the other approximations tend to underestimate the total energy.

%
%
%
%
%
%
%

%
%

\begin{figure}
\begin{center}
\includegraphics{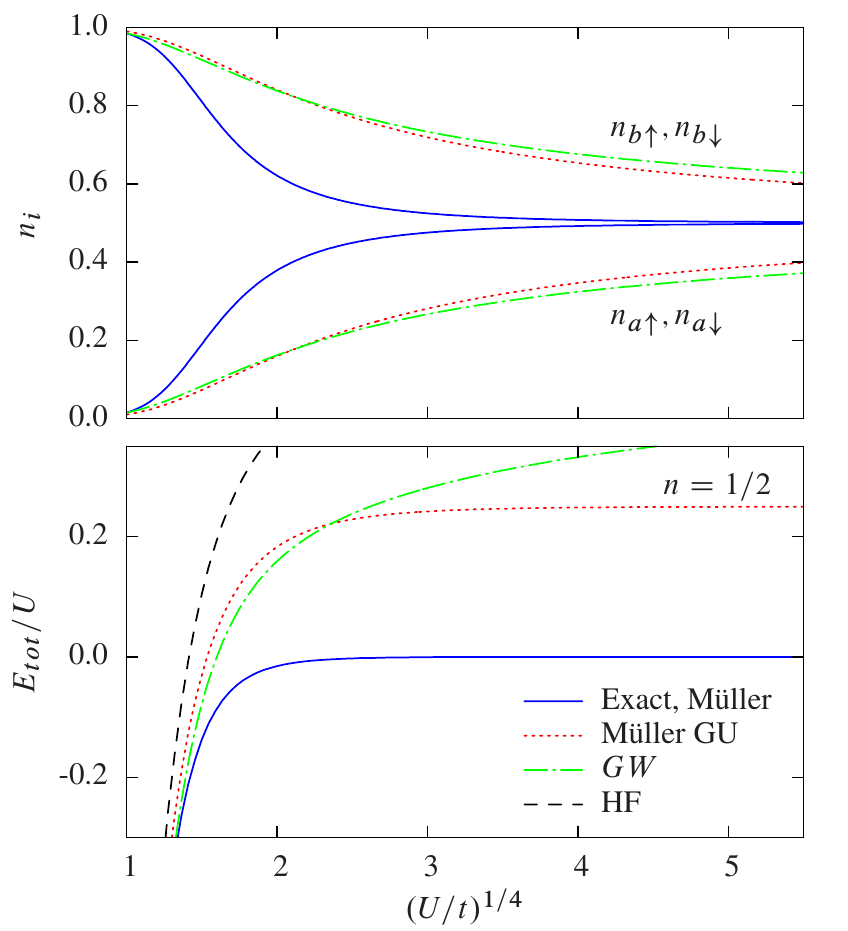}
\end{center}
\vspace{-15pt}
\caption{(Color online) Occupation numbers (upper panel) and total energy (lower panel) as function of $U/t$ at 1/2 filling: exact vs M\"uller functional w/wo self-interaction corrections (M\"uller and M\"uller GU, respectively) and $GW$. Total energies obtained using $GW$ Green's function in the Galitskii-Migdal formula (labelled $GW$) is also reported. } 
\label{Fig:figure2}
\end{figure} 

\subsubsection{1/2 filling}
The wavefunction at 1/2 filling reads 
$\ket{\Psi_0}=A\left(\ket{\uparrow\,\downarrow}-\ket{\downarrow\,\uparrow} \right)+ B\left(\ket{\uparrow\downarrow\,0}-\ket{0\,\downarrow\uparrow} \right)$ 
in the site basis, with $A=4t/[a(c-U)]$, $B=1/a$, $c=\sqrt{16t^2+u^2}$, and $a=\sqrt{2[16t^2/(c-U)^2+1]}$. When projected in the bonding/antibonding basis it reads $\ket{\Psi_0}=\sqrt{n_b}\ket{b\uparrow,b\downarrow}-\sqrt{n_a}\ket{a\uparrow,a\downarrow}$
with $\sqrt{n_b}=A+B$ and $\sqrt{n_a}=A-B$. Note that  $\ket{\Psi_0}$ depends on the square root of the occupation numbers; the success of the M\"uller functional with $\alpha=0.5$ at 1/2 filling is linked to this. \PRcomm{This functional is indeed closely related to the exact density-matrix functional for two-electron systems which is the L\"owdin-Shull 
functional \cite{Loewdin56}. The exact two-particle density 
matrix for such systems has an expansion in coefficients which are the 
square roots of the natural occupation numbers up to a sign \cite{ vanMeer14}. A proper selection 
of the signs (which in general is unknown) gives the 
exact result for two-electron systems.}  At $U/t=0$ the wavefunction is the single Slater determinant $\ket{\Psi_0}=\ket{b\uparrow,b\downarrow}$; increasing $U/t$ also the antibonding orbital becomes important, and eventually the full wavefunction becomes a linear combination of  the Slater determinants $\ket{b\uparrow,b\downarrow}$ and $\ket{a\uparrow,a\downarrow}$ with equal weight (see Fig.\ \ref{Fig:figure2}, upper panel). In terms of the site basis this means that for the noninteracting case each of the two electrons is equally distributed between the two sites, while increasing the interaction double occupancies become less probable. The optimal occupation numbers for the M\"uller functional are the exact ones. Clearly self-interaction corrections, as well as varying $\alpha$ in the range 0.5-1, spoil this result. Again $GW$ occupation numbers are very similar to self-interaction corrected M\"uller results. Note that $GW$ produces fractional occupation numbers with increasing $U/t$; eventually they go to 1/2 at $t=0$, but they go too slowly. 
\Pina{This means that at strong interaction $GW$ does not manage to well localize the two electrons each on one site, and spurious double occupancies are still present.}
%
%

Total energies are reported in the lower panel of Fig.\ \ref{Fig:figure2}. The total energy functional within the M\"uller approximation is not equal to the exact one, except at the exact occupation numbers. The \PRcomm{$GW$} result is similar to the result obtained using the self-interaction corrected M\"uller functional \PRcomm{below $(U/t)^{1/4} \simeq 3$; for stronger interaction they differ, but both overestimate the total energy. Our $GW$ result is in line with previous $GW$ calculations on the $H_2$ molecule \cite{Stan09,Caruso13} which show $GW$ to be very accurate close to equilibrium but to dramatically overestimate the total energy in the dissociation limit \footnote{Note that at 1/2 filling we use the particle-hole form of the Hubbard Hamiltonian \cite{romaniello_PRB12}. In the $GW$ removal/addition energies, the particle-hole symmetry is lost for $U\neq 0$ due to the lack of self-consistency;  we restore this symmetry by absorbing the static part of the self-energy ($U/2$) into the chemical potential. This alignment of the chemical potential corrects for the lack of self-consistency \cite{hedin65,Pollehn98}.}. Comparison with recent total energy calculations on the Hubbard dimer \cite{Olsen14} using the correlation energy expression obtained with the adiabatic-connection technique (see, e.g., \onlinecite{Ren12}) shows that RPA and beyond RPA approximations including excitonic effects give better result than $GW$ at strong interaction \footnote{Note that in Ref. \onlinecite{Ren12} the authors chose $\epsilon_0=0$ and $V_0=0$ in the Hubbard Hamiltonian. A non self-consistent $GW$ calculation in this case gives a total energy better than the RPA total energy; in our calculations, instead, by restoring the particle-hole symmetry \cite{romaniello_PRB12}, the $GW$ spectral properties are improved, but the total energy worsens.}.}

In conclusion, at one half filling the M\"uller functional gives the exact occupation numbers, and varying $\alpha$ towards 1 increases the discrepancy with this exact result.  $GW$ and M\"uller GU give similar occupation numbers, which merge with the exact ones at $t=0$, but at a lower speed than the exact values. This reflects the fact that these two approximations have difficulties to localize the two electrons one on each site, missing the atomic physics of strongly correlated electrons. Concerning the total energy both $GW$ and M\"uller GU  similarly overestimate the exact values, whereas the M\"uller functional gives the exact result. \PinaR{Increasing} $\alpha$ leads to higher total energies, with HF giving the worst agreement ($E_{tot}/t=-2+U/(2t)$).
\subsection{Removal/addition energies and Spectral function}
Exact removal and addition energies are reported in Figs \ref{Fig:figurePernal14}, \ref{Fig:figure3},  \ref{Fig:figure7}, and  \ref{Fig:figure5}. 
We analyze various ways to compute removal/addition energies within RDMFT, which elucidate the role played by an approximate exchange-correlation energy functional and by an approximate expression for the removal/addition energies. First we test the M\"uller-like approximations to the xc functional by combining the latter with the method proposed by Pernal and Cioslowski (EKT) \cite{Pernal200571} for the calculation of removal energies. The latter is based on the extended Koopman's theorem, which is an approximation; however for the Hubbard molecule at 1/4 and 1/2 filling the method gives the exact removal energies when combined with the exact exchange-correlation energy functional. This allows us to study the accuracy of the xc functional approximations. 
Second we test the DIF/DER method for the calculation of removal/addition energies by combining it with the exact xc functional. We then test the combination of the DIF/DER method and the M\"uller-like approximations to the xc energy functional.

Finally we combine the DIF/DER method and the M\"uller-like  functionals with the approximate expression for the spectral function given in Eq.\ (\ref{Eqn:DOS}). \PinaR{This is the approach used for the calculation of spectral functions of transition metal oxides and the results show a good agreement with experiment \cite{Sharma13}.} In the following we refer to this combination of approximations as the RDMFT spectral function.  Exact, $GW$, and RDMFT spectral functions are compared in  Figs \ref{Fig:figure4} and \ref{Fig:figure6}.

%
%
\subsubsection{1/4 filling}
At 1/4 filling the Hubbard molecule shows five quasiparticle energies (one removal (labelled $\omega_1$ in Fig.\ \ref{Fig:figurePernal14}) and 4 addition energies ($\omega_2$, $\omega_3$, $\omega_4$, $\omega_5$)) and one addition satellite energy ($\omega_6$). Satellites are weak removal or addition energies which acquire spectral weight with increasing interaction, \PinaR{whereas the intensity of quasiparticles decreases or remains constant.} If the exact energy functional is used (which, for one-electron, is just $E_{tot}=E_{kin}+E_{ext}$), then the EKT method produces the exact removal energy $\epsilon_0-t$ ($\omega_1$ in Fig. \ref{Fig:figurePernal14}). 
\PinaR{Using the M\"uller functional, instead, the EKT produces two removal energies (see result ``M\"uller+ EKT 1/4" in Fig.\ \ref{Fig:figurePernal14}). This is due to the fact that within this functional the antibonding occupation number $n_{a\uparrow}$ is not zero \footnote{Note that at $U=0$ $n_{a\uparrow}=0$ and the corresponding removal energy calculated with the EKT does not have a meaning. At $U=0$ the EKT yields only the exact removal energy.}}, and therefore more degrees of freedom are added to the problem. The energies do not match well with the exact results.
In the limit $U/t\rightarrow \infty$ the two \PinaR{removal} energies merge together at a value well off the exact one (see right panel of Fig.\ {\ref{Fig:figurePernal14}). 
In this limit the exact energies merge towards $\epsilon_0$ and $\epsilon_0+U$: this reflects the fact that in this limit the electron has equal probability to localize on one site or the other of the molecule, therefore one can have removal and addition energies (for a spin-down or a spin-up electron added to the empty site) at $\epsilon_0$ and an addition energy at $\epsilon_0+U$ (for a spin-down electron added to the site with one spin-up electron already). 
Improvements are obtained changing $\alpha$ from 0.5 to 1 (Hartree-Fock) as the exact functional is approached. Hartree-Fock, indeed, gives the exact total energy at 1/4 filling, due to an exact cancellation between Hartree and exchange energies.  If the lowest addition energy is calculated from the highest removal energy of the system at 1/2 filling the EKT yields the exact result (see result ``M\"uller+ EKT 1/2" in Fig.\ {\ref{Fig:figurePernal14}), but only because the M\"uller functional gives the exact total energy and occupation numbers at 1/2 filling ($N+1$-electron system).  

The DIF/DER method (Eqs (\ref{Eqn:DIF}) and (\ref{Eqn:DER})) performs as the EKT if the exact xc functional is used: it produces the exact removal energy and the exact second lowest addition energy. With the M\"uller functional it gives four energies (see result ``M\"uller+ DIF/DER" in Fig.\ {\ref{Fig:figure3}): only two energies are in good agreement with the exact ones ($\epsilon_{b\uparrow}$ and $\epsilon_{a\uparrow}$, calculated from $n_{b\uparrow}$ and $n_{a\uparrow}$, respectively), whereas for the other two ($\epsilon_{b\downarrow}$ and $\epsilon_{a\downarrow}$) we observe that each is approximately an average of two exact ones, namely $\epsilon_{b\downarrow}$ is an average of $\omega_2$ and $\omega_6$ \PinaR{\footnote{Note that at $U=0$ $\omega_6$ has zero spectral weight.}} and $\epsilon_{a\downarrow}$ of $\omega_{3}$ (or, equivalently, $\omega_4$) and $\omega_5$}. This can be understood considering that the $G_{b\uparrow}$ and $G_{a\uparrow}$ components of the one-body Green's functions have only one pole, whereas  $G_{b\downarrow}$ and $G_{a\downarrow}$ have two poles each; the corresponding occupation numbers (see Eq.\ (\ref{Eqn:N-G})) hence reflect these features. 

In general the spectral function profile is in overall good agreement with the exact one at moderately strong interaction $U/t$. For the spin-down channel $GW$ \PinaR{(right panel of Fig.\ \ref{Fig:figure4})} is slightly superior. \PinaR{It shows a very weak spurious satellite due to self-screening \cite{pina09} in the spin-up channel, but it correctly describes the spin-down satellite.}  
In the atomic limit ($t\rightarrow 0$) both DIF and DER methods show the same failure as $GW$:  for the spin-down spectral function the poles merge at $\epsilon_0+U/2$, unlike the exact result which shows a gap equal to $U$. We observe the same scenario increasing the number of sites \PinaR{(not shown)}. Self-interaction corrections to the M\"uller functional do not add any significant improvement to the picture. 
If the lowest addition energy (spin-down channel) is calculated from the highest removal energy of the $N+1$-electron system (1/2 filling), the method produces a gap, unlike the exact result, where the lowest addition energy  coincides with the highest removal energy ($\epsilon_0$). A similar error is found also in $GW$ and it is a consequence of the self-screening error $GW$ suffers from \cite{pina09}. 

\PinaR{In conclusion,} combining the approximate M\"uller-like functionals with the DIF/DER method \PinaR{significantly improves electron addition and removal energies seen as poles in the spectral function} with respect to the case where \PinaR{this functional} is used with the more advanced EKT. This indicates that there is  a cancellation of errors between the approximate M\"uller functional and the DIF/DER method, at least at 1/4 filling.


\begin{figure*}
\begin{center}
\includegraphics{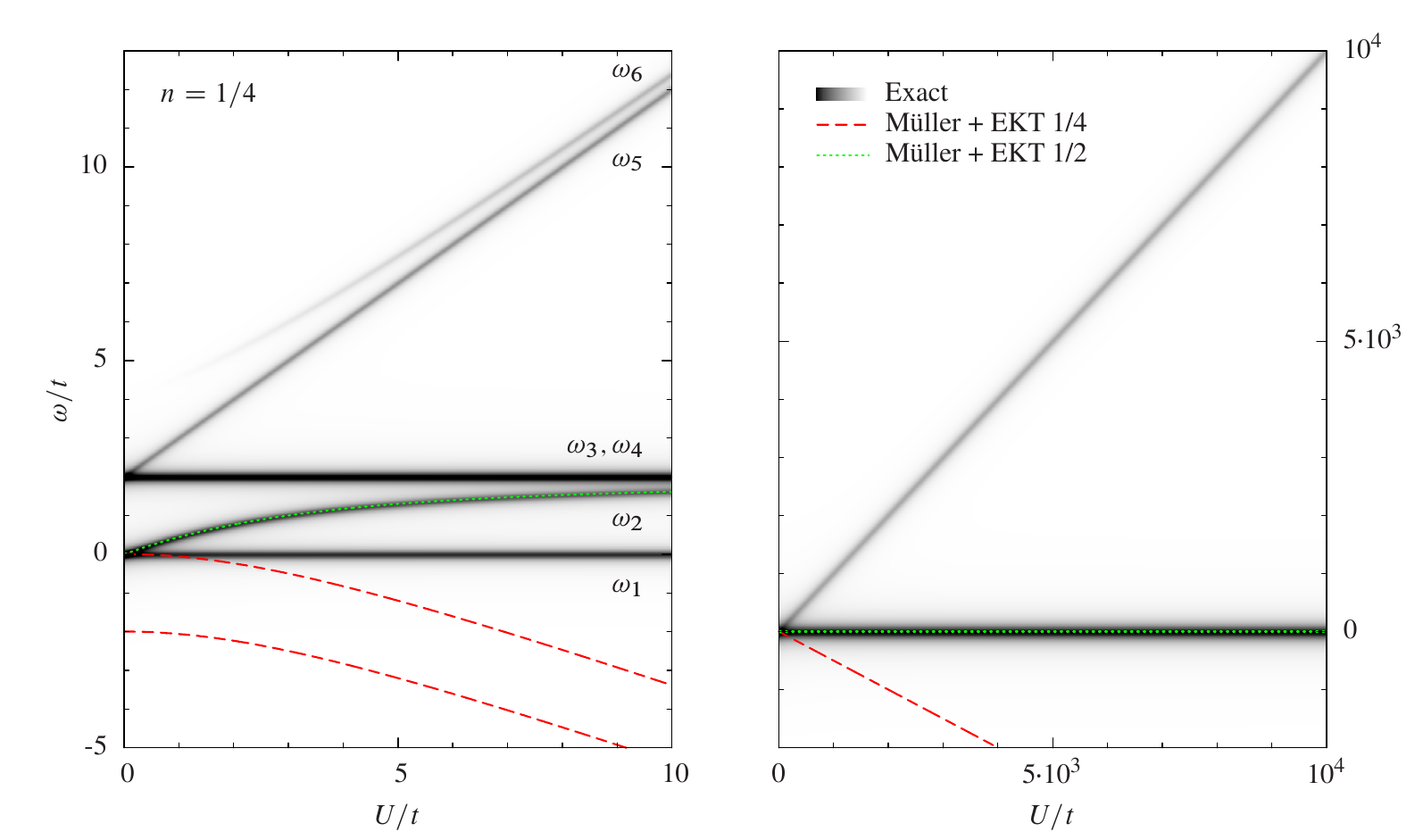} 
\end{center}
\vspace{-15pt}
\caption{(Color online) Removal and addition energies $\omega/t$ as function of $U/t$ at 1/4 filling: exact \textit{vs} EKT method used with the M\"uller functional \PinaR{(the label ``M\"uller+ EKT 1/4" refers to the removal energies, whereas ``M\"uller+ EKT 1/2" to the lowest addition energy calculated from the highest removal energy of the system at 1/2 filling)}. The labels $\omega_i$ indicate the exact energies. \PRcomm{The color gradient (from white to black) of the exact curves indicates increasing spectral weight; the energy $\omega_6$ is hence a satellite, since it has vanishing spectral weight at vanishing interaction.} \PinaR{The addition energy ``M\"uller+ EKT 1/2" is on top of the exact energy $\omega_2$, which goes to zero at strong interaction (right panel).}}
\label{Fig:figurePernal14}
\end{figure*}

\begin{figure*}
\begin{center}
\includegraphics{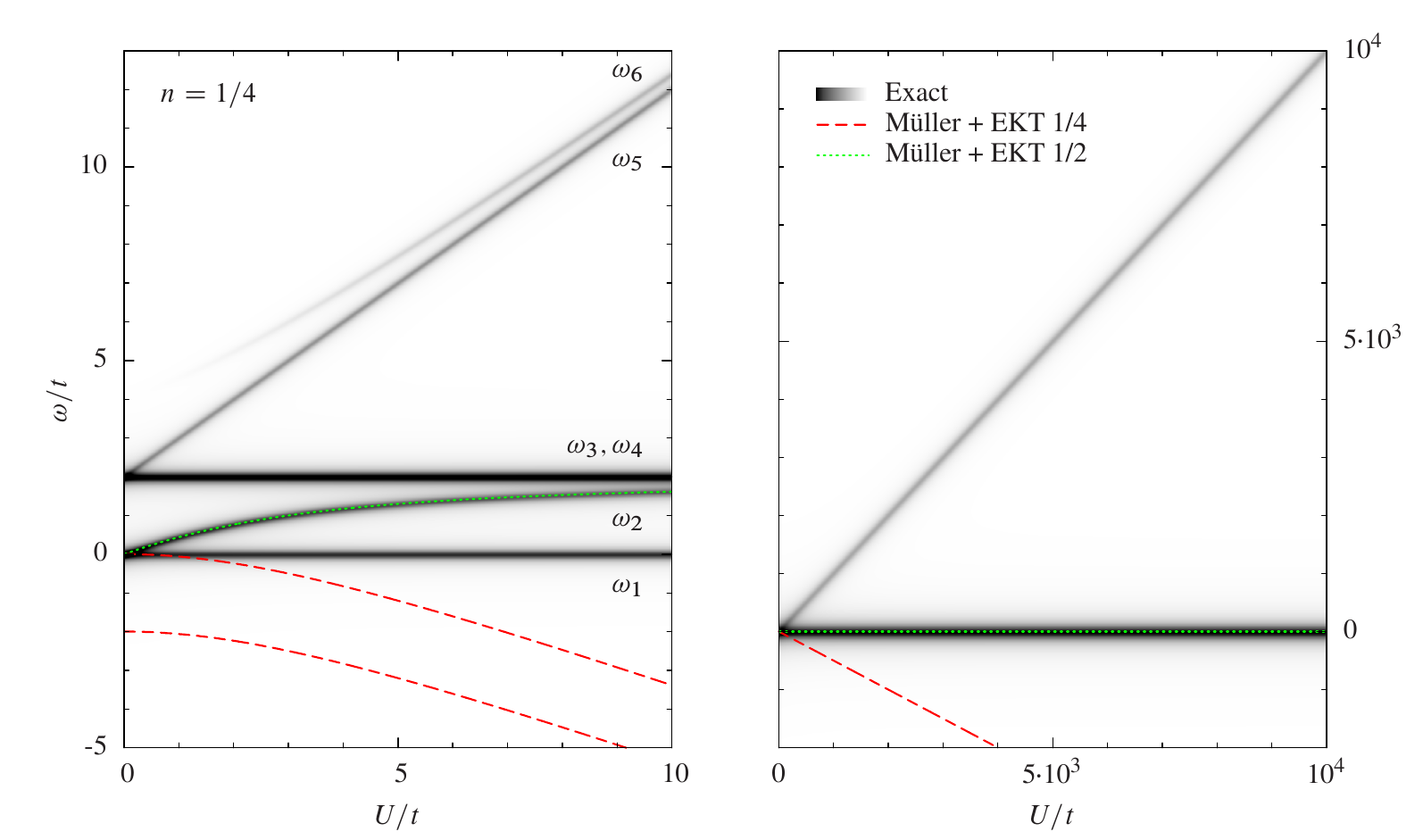} 
\end{center}
\vspace{-15pt}
\caption{(Color online) Removal and addition energies $\omega/t$ as function of $U/t$ at 1/4 filling: exact \textit{vs} DIF and DER methods used with the M\"uller functional. The labels $\epsilon_i$ indicate the bonding/antibonding energies obtained using the DIF/DER methods. \PRcomm{The color gradient of the exact curves has the same meaning as in Fig.\ref{Fig:figurePernal14}.} \PinaR{The energies $\epsilon_{b\downarrow}$ and $\epsilon_{a\downarrow}$ calculated with the DIF method are on top of those obtained with the DER method, and merge at $U/(2t)$ at strong interaction (right panel); the energies $\epsilon_{b\uparrow}$ and $\epsilon_{a\uparrow}$ calculated with the DER method merge to zero at strong interaction (right panel). }}\label{Fig:figure3}
\end{figure*}

\begin{figure*}
\begin{center}
\includegraphics{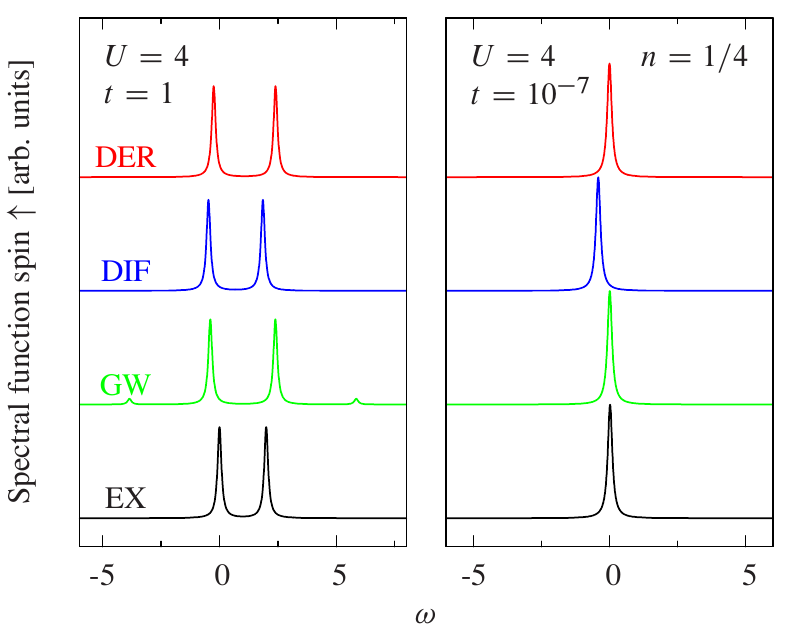}\hspace{12pt}
\includegraphics{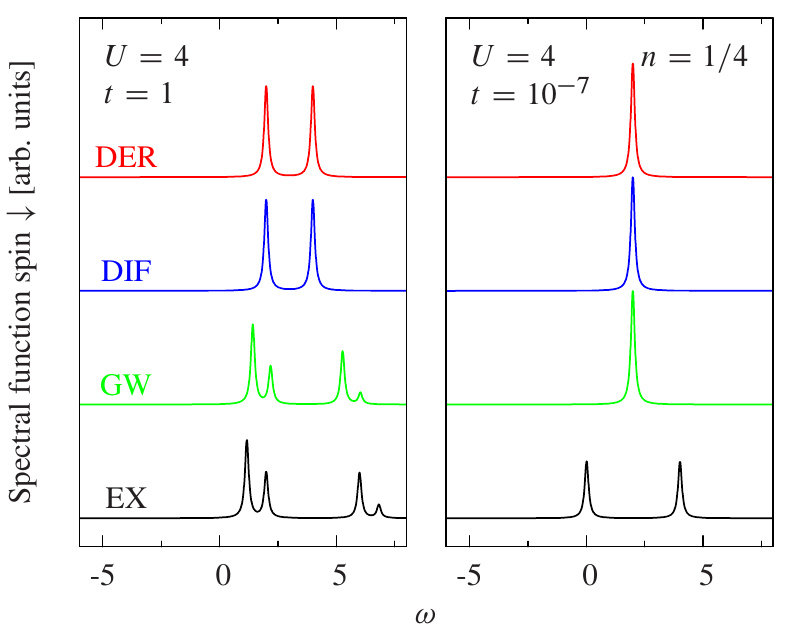}
\end{center}
\vspace{-15pt}
\caption{(Color online) Spectral function at 1/4 filling: exact \textit{vs} $GW$ and DER and DIF methods using M\"uller's functional.} \label{Fig:figure4}
\end{figure*}

\subsubsection{1/2 filling}
At 1/2 filling there are four quasiparticle energies (labeled $\omega_3$, $\omega_4$, $\omega_5$, $\omega_6$,  in Fig.\ \ref{Fig:figure7}) and four satellites ($\omega_1$, $\omega_2$, $\omega_7$, $\omega_8$). Using the M\"uller functional, which, in this case, gives the exact total energy at the exact occupation numbers, the EKT gives two doubly-degenerate energies: these are the exact removal energies, \PinaR{including a satellite.} To get the lowest addition energy one has to look at the $N+1$-electron system (3/4 filling); in this case the M\"uller functional does not reproduce the exact total energy and occupation numbers and, consequently, the EKT gives an addition energy that strongly departs from the exact one as $U/t$ increases.

Both DIF and DER methods give only two energies per spin channel, but their nature is in fact a mixture of quasiparticle and satellite energies \PinaR{and of electron addition and removal}; for example, we found that the energy $\epsilon_{b\uparrow}$ in Fig.\ \ref{Fig:figure7}) is roughly a weighted average of the satellite and quasiparticle energies $\omega_3$ and $\omega_7$, respectively. Again this can be understood by considering that the components $G_{b\sigma}$ and $G_{a\sigma}$ have two poles each, and therefore the corresponding occupation numbers reflect these features in the excitation energies $\epsilon_{b\sigma}$ and $\epsilon_{a\sigma}$. The results are quite different from the EKT, where one has both the removal quasiparticle and satellite energies. Since at 1/2 filling the M\"uller functional is exact, there is not the same cancellation of error as observed at 1/4 filling, and the DER/DIF method 
introduces, hence, quite a large error. 

In Fig.\ \ref{Fig:figure6} we report the exact spectral function vs the $GW$ and RDMFT spectral functions. As we see, only two peaks appear in the RDMFT spectra, which merge in the $t\rightarrow 0$ limit both for the DIF and DER method. Self-interaction corrections tend to open the gap,  but it is not enough \PinaR{in the strongly correlated dissociation limit.} 
Changing $\alpha$ does not improve the situation. DIF and DER, therefore, perform as bad as $GW$ in the atomic limit, whereas $GW$ is significantly superior at moderately strong interaction. We note that this conclusion is not restricted to the Hubbard molecule; we find the same scenario by increasing the number of sites up to 32. However, if the lowest addition energy is calculated from the highest removal energy of the $N+1$-electron system, the DIF/DER method yields a gap at best half of the exact one in the atomic limit. 

In conclusion, using an exact xc functional, the method of Ref.\ \onlinecite{Sharma13} has a large deviation from the exact results, both in the values and nature of the removal and addition energies. For moderately strong interaction $GW$ is clearly superior. In the atomic limit no gap is observed, like in $GW$, unless the $N+1$-electron system is considered for the calculation of the lowest addition energy.

\begin{figure*}
\begin{center}
\includegraphics{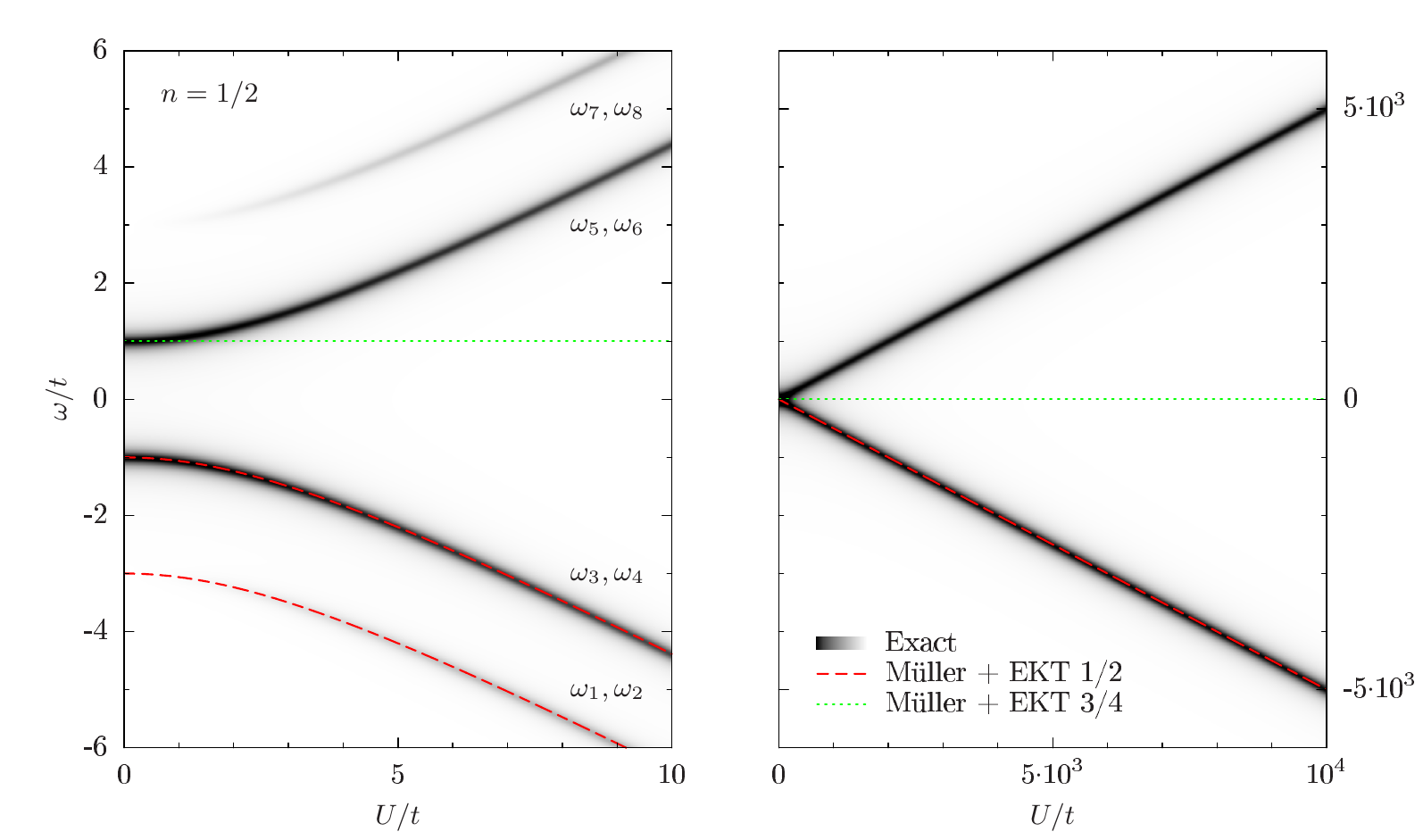} 
\end{center}
\vspace{-15pt}
\caption{(Color online) Removal and addition energies $\omega/t$ as function of $U/t$ at 1/2 filling: exact \textit{vs} EKT method used with the M\"uller functional \PinaR{(the label ``M\"uller+ EKT 1/2" refers to the removal energies, whereas ``M\"uller+ EKT 3/4" to the lowest addition energy calculated from the highest removal energy of the system at 3/4 filling)}. The labels $\omega_i$ indicate the exact energies. \PRcomm{The color gradient (from white to black) of the exact curves indicates increasing spectral weight; the energies $\omega_1,\,\omega_2,\,\omega_7,\,\omega_8$ are hence satellites, since they have vanishing spectral weight at vanishing interaction.}}\label{Fig:figure7}
\end{figure*}

\begin{figure*}
\begin{center}
\includegraphics{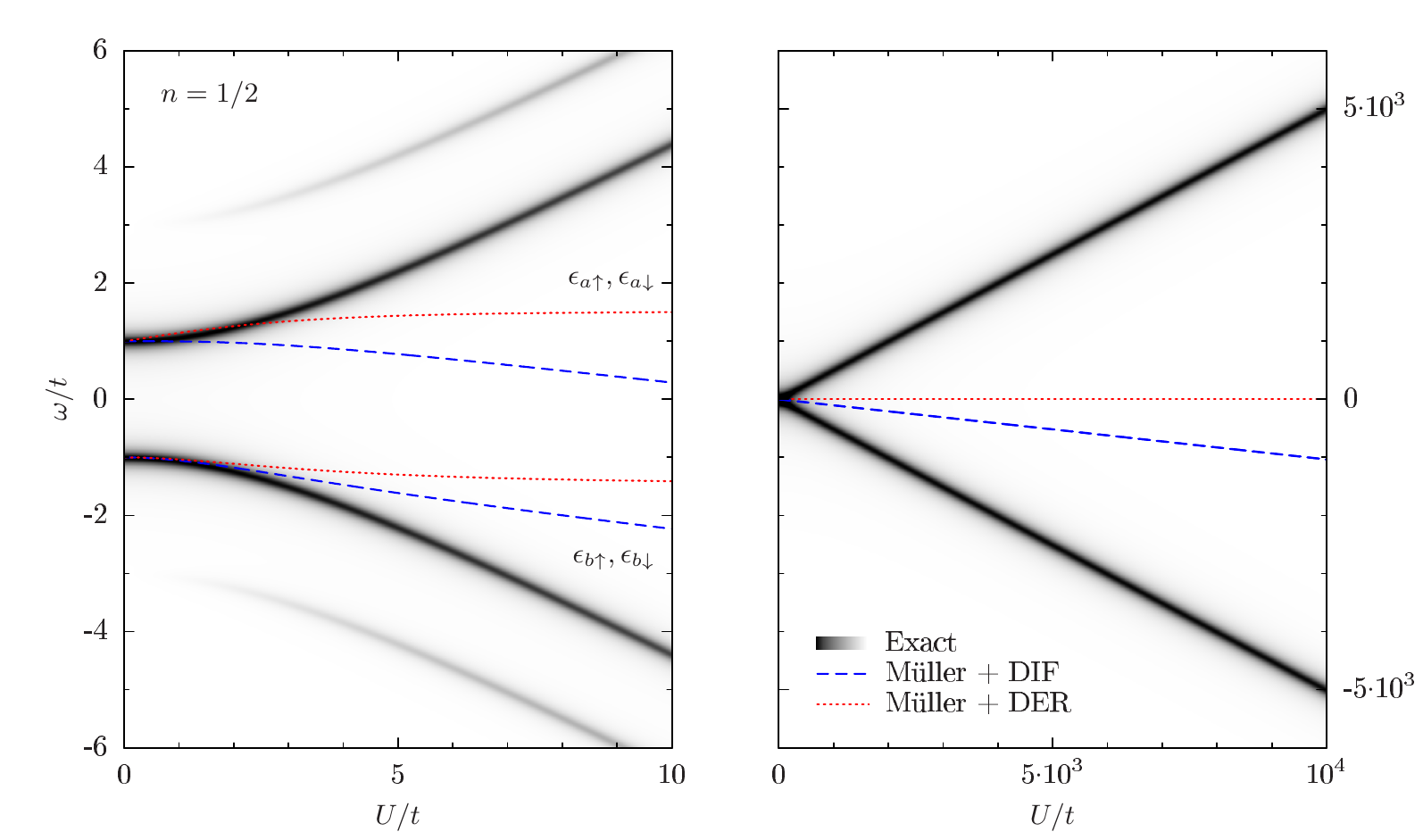}  
\end{center}
\vspace{-15pt}
\caption{(Color online) Removal and addition energies $\omega/t$ as function of $U/t$ at 1/2 filling: exact \textit{vs} DIF and DER methods used with the M\"uller functional. The labels $\epsilon_i$ indicate the bonding/antibonding energies obtained using the DIF/DER methods.  \PRcomm{The color gradient of the exact curves has the same meaning as in Fig.\ref{Fig:figure7}}.}\label{Fig:figure5}
\end{figure*}



\begin{figure}
\begin{center}
\includegraphics{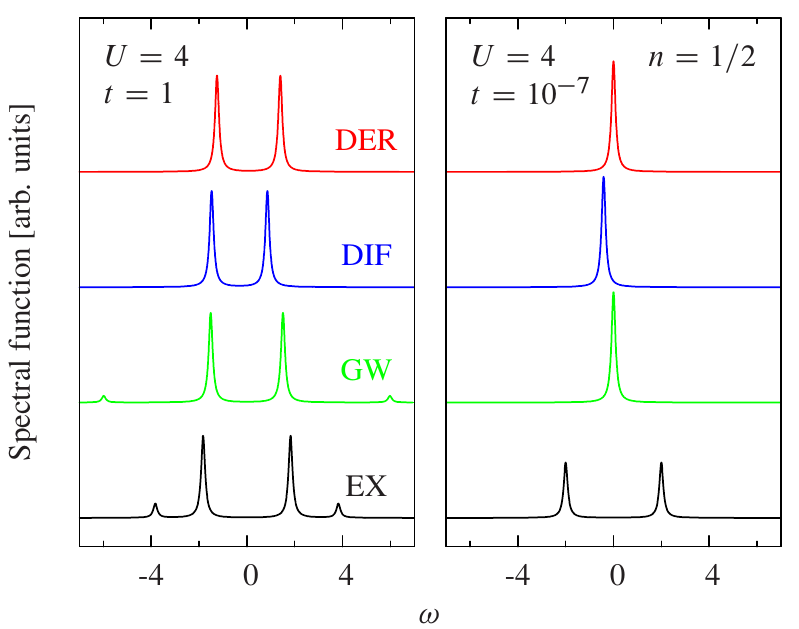} 
\end{center}
\vspace{-15pt}
\caption{(Color online) Spectral function at 1/2 filling: exact \textit{vs} $GW$ and DER and DIF methods using M\"uller's functional.}\label{Fig:figure6}
\end{figure}

\subsection{Occupation numbers and correlation}
Occupation numbers are an \Pina{indicator} of correlation. 
\Pina{However, in general, a physical system is subjected to measurements, and measurements change the system. One hence cannot look only at the occupation numbers \PinaR{of the initial system to understand observed correlation effects}. For example, }in Sec.\ \ref{Sec:TOT_E_oneforth} we looked at the Hubbard molecule at 1/4 filling for which the exact occupation numbers are either zero or one. This is a clear example in which the occupation numbers indicate no correlation in the system, but there are correlation effects when looking at the system using measurements of electron addition. \PRcomm{In this case, indeed, the spectrum shows, besides quasiparticle peaks, also a satellite ($\omega_6$ in Fig.\ \ref{Fig:figurePernal14} ), which is \Pina{a pure signature of correlation}.}  
Therefore whether a system ``is" correlated or not depends on how one looks at it.

\PinaR{Let us examine this point further,} by looking at the Hubbard molecule at 1/2 filling as example. In the atomic limit the spin-singlet ground state $\ket{\Psi_0}=1/\sqrt{2}[\ket{\uparrow\,\downarrow}-\ket{\downarrow\,\uparrow}]$ becomes degenerate with the spin-symmetry broken state $\ket{\Psi_0}=\ket{\uparrow\,\downarrow}$ (or, equivalently, $\ket{\Psi_0}=\ket{\downarrow\,\uparrow}$)\PinaR{, which is also an eigenstate of the system in this limit.} We note that in the spin-symmetry broken case both $GW$ and RDMFT give the exact result for total energy, occupation numbers, and spectral function.} 

Let us first focus on the spectral function. 
From Fig. \ref{Fig:figure6} we see that the exact spectral function of the spin singlet  (lowest right panel) shows two peaks, one at $\epsilon_0$ and one at $\epsilon_0+U$, both for spin-up and spin-down channels. For the \PRcomm{spin-symmetry broken} case the components of the one-body Green's functions show the spin-symmetry breaking nature of the ground state, i.e. $G_{ii\uparrow}\neq G_{ii\downarrow}$ and $G_{11\uparrow}=G_{22\downarrow}$, $G_{11\downarrow}=G_{22\uparrow}$, and hence are different from the ones of the singlet case (all diagonal components the same, all off diagonal components the same). However the spin-resolved total spectral function, i.e. $A_{\sigma}(\omega)=\sum_i \left|\Im G_{ii\sigma}(\omega)\right|/\pi$, is the same for the two spin structures: one can remove a spin-up or spin-down electron with energy $\epsilon_0$, and can add a spin-up or spin-down electron with energy $\epsilon_0+U$. \PinaR{This corresponds to the fact that in real systems photoemission experiments find similar results for the symmetric and symmetry-broken phases}, although there are \PinaR{of course small} differences introduced by changes in geometry, density etc. 
\PinaR{For example, in NiO no significant changes in the valence band structure are detected passing from the paramagnetic to the antiferromagnetc phase \cite{Tjernberg_PRB96}.}

What about the occupation numbers? \Pina{For the spin-singlet structure the natural orbitals are the bonding/antibonding orbitals and the occupation numbers are $n_{b\uparrow}=n_{b\downarrow}=1/2$ and $n_{a\uparrow}=n_{a\downarrow}=1/2$ (see Fig. \ref{Fig:figure2}). For the spin-symmetry broken structure, characterized by a single Slater determinant, the natural orbitals are the site orbitals $\psi_{1\sigma/2\sigma}=\left(\phi_{b\sigma}\pm\phi_{a\sigma}\right)/\sqrt{2}$ with occupation numbers $n_{1\uparrow}=n_{2\downarrow}=1$ and $n_{1\downarrow}=n_{2\uparrow}=0$. Spin-resolved occupation numbers are, hence, different for the two spin structures. \PinaR{It is now interesting to examine whether this difference could be measured. 

The most direct experimental route to access occupation numbers is Compton scattering.}} The Compton profile gives information about the momentum distribution, i.e. the probability to observe a particle of momentum $\mathbf{p}$ (see, e.g. Refs \onlinecite{olevano2010,olevano2012}). This can be expressed in terms of the Fourier transform in momentum space of the density matrix, as
\begin{eqnarray}
n(\mathbf{p})\propto\int d\mathbf{r}d\mathbf{r'}e^{-i\mathbf{p}\cdot(\mathbf{r}-\mathbf{r'})}\gamma(\mathbf{r},\mathbf{r'}),
\label{Eqn:momentum_D}
\end{eqnarray}
where we defined $\gamma(\mathbf{r},\mathbf{r'})=\sum_{\sigma}\PinaR{\sum_{ss'}}\chi^*_{\sigma}(s)\gamma(\mathbf{x},\mathbf{x'})\chi_{\sigma}(s')$, with $\chi_{\sigma}(s)$ the spin function, \PRcomm{which is defined as $\chi_{\uparrow}(1/2)=\chi_{\downarrow}(-1/2)=1$ and  $\chi_{\uparrow}(-1/2)=\chi_{\downarrow}(1/2)=0$}. The Fourier transform (\ref{Eqn:momentum_D}) gives the matrix elements of the density matrix in a basis of plane waves $\phi_{p\sigma}(\mathbf{r},s)=1/\sqrt{\Omega} \, e^{i\mathbf{p}\cdot\mathbf{r}}\chi_\sigma(s)$, which are the exact one-electron eigenstates of the free electron gas, i.e. the perfectly translationally invariant system. \PinaR{The question is hence what one could observe looking at matrix elements of the density matrix in the symmetry basis.} \PinaR{The analogous basis for the Hubbard molecule which reflects the symmetry of the system is the bonding/antibonding basis $\{\phi_{b\sigma/a\sigma}\}$.} \Pina{Since this basis is the basis of natural orbitals for the spin-singlet system, in this case the \PinaR{analog of the} Compton profile \PinaR{gives} the occupation numbers $n_b=1$ (with $n_{b\uparrow}=n_{b\downarrow}=1/2$) and $n_a=1$ (with $n_{a\uparrow}=n_{a\downarrow}=1/2$).
 One gets the same \PinaR{result} for the spin-symmetry broken structure. Note, however, that in this case, unlike for the spin-singlet structure, \PinaR{this distribution corresponds to  density-matrix elements that are not occupation numbers}. This is because the bonding/antibondig basis in which the density matrix is projected is not the basis of natural orbitals for the spin-symmetry broken structure.}
%
In fact not even a spin-resolved ``Compton profile" would distinguish between the two spin structures, since for the spin-broken symmetry structure \Pina{one gets the density-matrix elements} $n_{b\sigma/a\sigma}=\int d\mathbf{x}d\mathbf{x}'\phi^*_{b\sigma/a\sigma}(\mathbf{x})\left[\sum_{i=1\uparrow,2\downarrow}\psi_i(\mathbf{x})\psi_i^*(\mathbf{x}')\right]\phi_{b\sigma/a\sigma}(\mathbf{x}')=1/2$, as for the spin-singlet case. To distinguish between the two cases one should \Pina{measure other aspects of the density matrix\PinaR{, for example, carry out} a spin- and space-resolved measurement of the density matrix elements.}  In this case the density matrix is projected in the site basis, which gives \Pina{the density matrix elements} $n_{1\uparrow}=n_{2\uparrow}=n_{1\downarrow}=n_{2\downarrow}=\frac{1}{2}$ for the spin-singlet and \Pina{the occupation numbers} $n_{1\uparrow}=n_{2\downarrow}=1$, $n_{1\downarrow}=n_{2\uparrow}=0$ for the spin-symmetry broken case.
}


 %
%
\section{Conclusions\label{Sec:Conclusions}}
We analyzed the results for total energy, natural occupation numbers, removal/addition energies, and spectral function for the Hubbard molecule at 1/4 and 1/2 filling by using reduced density-matrix functional theory and many-body perturbation theory within standard approximations to electron correlation, namely M\"uller-like functionals and $GW$, respectively. In general there is no M\"uller-like functional which works well at both one fourth and half filling: for the former the Hartree-Fock functional gives the exact total energy and occupation numbers, whereas for the latter the original M\"uller functional does the job. Other M\"uller-like functionals underestimate the total energy at $1/4$ filling and overestimate it at $1/2$ filling, like $GW$. The same behavior is found for the occupation numbers, which deviate in a similar way as $GW$ from the exact results.

We also analyzed various approximate methods to obtain removal/addition energies and spectral functions from RDMFT. Our results for the removal/addition energies and spectral function obtained using the approximate method of Ref.\ \onlinecite{Sharma13} suggest a cancellation of errors between the latter and M\"uller-like approximations to electron correlations. Moreover the spectral peaks can have a mixed removal and addition nature as well as a mixed quasiparticle and satellite nature. At moderately strong interaction, $GW$ is superior. In the strongly correlated electron regime, which is obtained by stretching the molecule (atomic limit), we found that both RDMFT and $GW$ fail for a spin-singlet ground state, whereas they give the exact results for the spin-symmetry-broken case. \PinaR{Because the Hubbard molecule is a simple test case, it shines light on the content, successes and limits of current RDMFT approaches and we believe that arguments like those based on symmetry and symmetry breaking can be safely generalized to improve our understanding of real systems.}

\begin{acknowledgments}
The research leading to these results has received funding
from the European Research Council under the European
Union's Seventh Framework Programme (FP/2007-
2013) / ERC Grant Agreement n. 320971.
Discussion within the Collaboration Team on Correlation of the European Theoretical Spectroscopy facility (ETSF) is greatly acknowledged. S. DiS. and P. R. thank S. Sharma, N. N. Lathiotakis, and F. G. Eich for fruitful discussions.  
\end{acknowledgments}
\appendix
\section{Bonding-antibonding basis}\label{app_bondant}
The bonding/antibonding basis $\{\phi_i\}$ is defined as
\begin{eqnarray}
\phi_{b\sigma}&=&\frac{1}{\sqrt{2}}\psi_{1\sigma}+\frac{1}{\sqrt{2}}\psi_{2\sigma}\label{Eqn:B}\\
\phi_{a\sigma}&=&\frac{1}{\sqrt{2}}\psi_{1\sigma}-\frac{1}{\sqrt{2}}\psi_{2\sigma}\label{Eqn:A},
\end{eqnarray}
where $\{\psi_i\}$ is the site basis. 

In Ref \onlinecite{pina09} the results of the Hubbard molecule 1/4 and 1/2 filling are given using a basis of Slater determinants $|...\rangle$ constructed using the site basis. In tables \ref{table1} and \ref{table2} we gives the transformation of this Slater determinants from the site basis to the bonding/antibonding basis. 
\subsection{1/4 filling}
\begin{table}
\begin{center}
    \begin{tabular}{| l | c c c c |}
    \hline
                         & $\ket{b\uparrow}$ & $\ket{b\downarrow}$ & $\ket{a\uparrow}$ & $\ket{a\downarrow}$\\ \hline
    $\ket{\uparrow,0}$   &      $1/\sqrt{2}$ &             0  &      $1/\sqrt{2}$ &                  0 \\
    $\ket{\downarrow,0}$ &                 0 & $1/\sqrt{2}$   & 0             & $1/\sqrt{2}$              \\
    $\ket{0,\uparrow}$   &      $1/\sqrt{2}$ & 0              & $-1/\sqrt{2}$ & 0  \\
    $\ket{0,\downarrow}$ &                 0 & $1/\sqrt{2}$   & 0             & $-1/\sqrt{2}$              \\
    \hline
    \end{tabular}
    \caption{Coefficients of the transformation from site to bonding/antibonding basis for 1 electron.}
   \label{table1}
   \end{center}
\end{table}

The eigenstates of the Hamiltonian at 1/4 filling are
$\ket{b\uparrow}$ (the ground state), $\ket{b\downarrow}$, $\ket{a\uparrow}$, $\ket{a\downarrow}$
with energies $\epsilon_0-t$, $\epsilon_0-t$, $\epsilon_0+t$, $\epsilon_0+t$.

\subsection{1/2 filling}
\begin{table}
\begin{center}
    \begin{tabular}{| l | l l l l l l |}
    \hline
     & $\ket{b\uparrow,b\downarrow}$ & $\ket{b\uparrow,a\uparrow}$ & $\ket{b\uparrow,a\downarrow}$ & $\ket{b\downarrow,a\uparrow}$ & $\ket{b\downarrow,a\downarrow}$   & $\ket{a\uparrow,a\downarrow}$\\ \hline
    $\ket{\uparrow,\downarrow}$   &  1/2 & 0  & -1/2 & -1/2 & 0  & -1/2 \\
    $\ket{\downarrow,\uparrow}$   & -1/2 & 0  & -1/2 & -1/2 & 0  &  1/2 \\
    $\ket{\uparrow,\uparrow}$     &   0  & -1 & 0 & 0 & 0   & 0 \\
    $\ket{\downarrow,\downarrow}$ &   0  & 0  & 0 & 0 & -1  & 0 \\
    $\ket{\uparrow\downarrow,0}$  &  1/2 & 0  & 1/2 & -1/2  & 0  & 1/2 \\
    $\ket{0,\uparrow\downarrow}$  &  1/2 & 0  & -1/2   &    1/2  & 0 & 1/2 \\
    \hline
    \end{tabular}
    \caption{Coefficients of the transformation from site to bonding/antibonding basis for 2 electrons.}
 \label{table2}
\end{center}
\end{table}
The eigenstates of the hamiltonian at 1/2 filling in the bonding/antibonding rappresentation are given in table \ref{table3}. Here $c^2=16t^2+U^2$ and $a^2=2\left[16t^2/(c-U)^2+1\right]$, and $b^2=2\left[16t^2/(c+U)^2+1\right]$.

\begin{table*}
\begin{center}
    \begin{tabular}{| l | l l l l l l |}
    \hline
    $E_i$ & $\ket{b\uparrow,b\downarrow}$ & $\ket{b\uparrow,a\uparrow}$ & $\ket{b\uparrow,a\downarrow}$ & $\ket{b\downarrow,a\uparrow}$ & $\ket{b\downarrow,a\downarrow}$ & $\ket{a\uparrow,a\downarrow}$\\ \hline
    $2\epsilon_0+(u-c)/2$ & $(1+4t/(c-u))/a$ & 0  & 0             & 0             & 0  & $(1-4t/(c-u))/a$ \\
    $2\epsilon_0+(u+c)/2$ & $(1-4t/(c+u))/b$ & 0  & 0             & 0             & 0  & $(1+4t/(c+u))/b$ \\
    $2\epsilon_0+u$       &               0  & 0  & $-1/\sqrt{2}$ & $1/\sqrt{2}$  & 0  & 0 \\
    $2\epsilon_0$         &               0  & 0  & 0             & 0             & -1 & 0 \\
    $2\epsilon_0$         &               0  & -1 & 0             & 0             & 0  & 0 \\
    $2\epsilon_0$         &               0  & 0  & $-1/\sqrt{2}$ & $-1/\sqrt{2}$ & 0  & 0 \\
    \hline
    \end{tabular}
    \caption{Eigenvalues and coefficients for the two electron system in bonding/antibonding basis.}
    \label{table3}
\end{center}
\end{table*}


%

\bibliographystyle{apsrev}
 \newcommand{\noop}[1]{}

\end{document}